\documentclass[%
 reprint,
 amsmath,amssymb,
 aps,
]{revtex4-2}

\usepackage{graphicx}% Include figure files
\usepackage{dcolumn}% Align table columns on decimal point
\usepackage{bm}% bold math
\usepackage{hyperref}
\usepackage[capitalise]{cleveref}
\usepackage{ amssymb }

\usepackage{pgf}
\usepackage{layouts}
\usepackage{todonotes}

\def\equationautorefname{Eq.}
\def\figureautorefname{Fig.}
\newcommand{\Autoref}[1]{%
  \begingroup%
  \renewcommand\equationautorefname{Equation}%
  \renewcommand\figureautorefname{Figure}%
  \autoref{#1}%
  \endgroup%
  }

\begin{document}
\newcommand{\e}[1]{\times 10^{#1}}

\preprint{APS/123-QED}

\title{Improved limits on the coupling of ultralight bosonic dark matter to photons from optical atomic clock comparisons}% Force line breaks with \\

\author{M.~Filzinger}
%\affiliation{Physikalisch-Technische Bundesanstalt, Bundesallee 100, 38116 Braunschweig, Germany}
\author{S.~D\"orscher}
%\affiliation{Physikalisch-Technische Bundesanstalt, Bundesallee 100, 38116 Braunschweig, Germany}
\author{R.~Lange}
%\affiliation{Physikalisch-Technische Bundesanstalt, Bundesallee 100, 38116 Braunschweig, Germany}
\author{J.~Klose}
%\affiliation{Physikalisch-Technische Bundesanstalt, Bundesallee 100, 38116 Braunschweig, Germany}
\author{M.~Steinel}
%\affiliation{Physikalisch-Technische Bundesanstalt, Bundesallee 100, 38116 Braunschweig, Germany}
\author{E.~Benkler}
%\affiliation{Physikalisch-Technische Bundesanstalt, Bundesallee 100, 38116 Braunschweig, Germany}
\author{E.~Peik}
%\affiliation{Physikalisch-Technische Bundesanstalt, Bundesallee 100, 38116 Braunschweig, Germany}
\author{C.~Lisdat}
%\affiliation{Physikalisch-Technische Bundesanstalt, Bundesallee 100, 38116 Braunschweig, Germany}
\author{N.~Huntemann}
\email[]{nils.huntemann@ptb.de}
\affiliation{Physikalisch-Technische Bundesanstalt, Bundesallee 100, 38116 Braunschweig, Germany}

\date{\today}% It is always \today, today,

\begin{abstract}
We present improved constraints on the coupling of ultralight bosonic dark matter to photons based on long-term measurements of two optical frequency ratios. In these optical clock comparisons, we relate the frequency of the ${}^2S_{1/2} (F=0)\leftrightarrow {}^2F_{7/2} (F=3)$ electric-octupole (E3) transition in \textsuperscript{171}Yb\textsuperscript{+} to that of the ${}^2S_{1/2} (F=0)\leftrightarrow \,{}^2D_{3/2} (F=2)$ electric-quadrupole (E2) transition of the same ion, and to that of the ${}^1S_0\leftrightarrow\,{}^3P_0$ transition in \textsuperscript{87}Sr. Measurements of the first frequency ratio $\nu_\textrm{E3}/\nu_\textrm{E2}$ are performed via interleaved interrogation of both transitions in a single ion. The comparison of the single-ion clock based on the E3 transition with a strontium optical lattice clock yields the second frequency ratio $\nu_\textrm{E3}/\nu_\textrm{Sr}$. By constraining  oscillations of the fine-structure constant $\alpha$ with these measurement results, we improve existing bounds on the scalar coupling $d_e$ of ultralight dark matter to photons for dark matter masses in the range of about $ 10^{-24}-10^{-17}\,\textrm{eV}/c^2$. These results constitute an improvement by more than an order of magnitude over previous investigations for most of this range. We also use the repeated measurements of $\nu_\textrm{E3}/\nu_\textrm{E2}$ to improve existing limits on a linear temporal drift of $\alpha$ and its coupling to gravity.
\end{abstract}
\maketitle
Even though dark matter makes up the majority of the matter in our universe, its microscopic properties and non-gravitational interactions are still a mystery. One well-motivated dark matter model is that of ultralight bosons (see \cite{Kimball2022, Antypas2022} for recent reviews). Bosons with a mass $m_\varphi$ well below $ 1\,\textrm{eV}/c^2$, with $c$ the speed of light, are expected to behave like a classical coherent wave, with an oscillation frequency given by their Compton frequency $\omega= 2 \pi\, m_{\varphi}c^2/ h $, where $h$ is the Plack constant, and a finite coherence time $\tau_{\textrm{coh}}\approx h/m_\varphi \Delta v^2$ due to the velocity spread $\Delta v\approx 10^{-3} c$ of dark matter in our galaxy. 

The interaction of this ultralight bosonic dark matter (UBDM) with standard model particles is expected to lead to corresponding oscillations in fundamental constants, such as the fine-structure constant, the electron mass, and the quantum chromodynamics mass scale \cite{Arvanitaki2015}. Experiments based on atomic and optical techniques can be sensitive to changes of these parameters and have provided some of the most stringent limits on the couplings of UBDM to standard model particles to date \cite{VanTilburg2015,Hees2016,Kennedy2020,Beloy2021,Savalle2021,Tretiak2022, Kobayashi2022,Oswald2022}. 

In this work, we are concerned with a coupling of the dimensionless dark matter field $\varphi$ to photons contributing a term to the Lagrangian density $\mathcal{L}$
\begin{equation}
    \mathcal{L}\supset \varphi\frac{d_e}{4\mu_0}\, F_{\mu\nu}F^{\mu\nu}\,\textrm{,}\label{eq:Lagr}
\end{equation}
with coupling constant $d_e$, $F_{\mu\nu}$ the electromagnetic field tensor, and $\mu_0$ the vacuum permeability. This coupling leads to oscillations of the fine-structure constant \cite{Arvanitaki2015}:
\begin{equation}
    \alpha(t)\approx\alpha \, \left[1+d_e \varphi_0 \cos(\omega t+\delta)\right]\label{eq:alpha}
\end{equation}
with $\delta$ an unknown phase, and  the dimensionless amplitude
\begin{equation}
  \varphi_0\equiv\sqrt{\frac{4\pi G\hbar^2}{c^6}}\, \frac{\sqrt{2\rho_{DM}}}{m_{\varphi}}\,\textrm{,}\label{eq:amp}
\end{equation}
where $\rho_{DM}$ denotes the dark matter energy density and $G$ the gravitational constant.

Since the energy of atomic levels depends on the fine structure constant $\alpha$, atomic transition frequencies can be sensitive probes to changes of its value. In particular, a sinusoidal oscillation of $\alpha$ due to a coupling of UBDM to photons can lead to a magnified oscillation in the ratio of two optical atomic frequencies $\nu_{1}$ and $\nu_{2}$
\begin{equation}
\frac{\Delta (\nu_{1}/\nu_{2})}{\nu_{1}/\nu_{2}} = - k_{\alpha}\frac{\Delta \alpha}{\alpha}\,\textrm{,}\label{eq:sens}
\end{equation}
if the magnitude of their differential sensitivity $k_\alpha$ to changes of $\alpha$ is larger than one.

In this letter, we present a search for such sinusoidal modulations in our measurements of two optical frequency ratios. The excited state of the ${}^2S_{1/2} (F=0)\leftrightarrow\,{}^2F_{7/2} (F=3)$ electric-octupole (E3) transition in \textsuperscript{171}Yb\textsuperscript{+} features a single hole in the otherwise filled $4f$ shell, while the ground state is characterized by a single valence electron $4f^{14} 6s^1$. The proximity of the $4f$ shell to the nucleus of this heavy ion yields an intuitive explanation for large relativistic contributions to the E3 excited state energy. This makes optical clocks based on the E3 transition the most sensitive to variations of $\alpha$ presently in operation. We utilize this high sensitivity by comparing the E3 transition frequency $\nu_\textrm{E3}$ to two other transition frequencies: The ${}^2S_{1/2} (F=0)\leftrightarrow\, {}^2D_{3/2} (F=2)$ electric-quadrupole (E2) transition frequency $\nu_\textrm{E2}$ of the same ion and the ${}^1S_0\leftrightarrow\,{}^3P_0$ transition frequency $\nu_\textrm{Sr}$ in \textsuperscript{87}Sr. The sensitivity $k_{\alpha}$ has been calculated as $6.95$ for $\nu_\textrm{E3}/\nu_\textrm{E2}$ and $6.01$ for $\nu_\textrm{E3}/\nu_\textrm{Sr}$ \cite{Flambaum2009}. The strong $\alpha$-dependence in these frequency comparisons enables us to search for a coupling $d_e$ of UBDM to photons with high sensitivity.

The single-ion clock used for the measurements reported here  has previously been evaluated to a fractional uncertainty of $2.7\times 10^{-18}$ on the E3 transition \cite{sanner2019a}, and $33\times 10^{-18}$ on the E2 transition \cite{Lange2021}. Measurements of the optical frequency ratio $\nu_\textrm{E3}/\nu_\textrm{E2}$ have been performed at PTB since 2016 \cite{Lange2021}. The data reported previously was predominantly obtained by comparing two single-ion clocks, where one was realizing $\nu_\textrm{E3}$ and the other one $\nu_\textrm{E2}$. Since autumn 2020, both frequencies have been realized with the same apparatus via interleaved interrogation, i.e. the single ion is probed on both transitions in an alternating fashion. The E3 transition is interrogated using Rabi-controlled hyper-Ramsey spectroscopy~\cite{Huntemann2016} with a dark time of $500\,$ms, while we use standard Rabi interrogation with $42\,$ms long pulses for the E2 transition, because of its short $53\,$ms excited-state lifetime~\cite{Yu2000}. 
 The E2 transition frequency is averaged over three mutually perpendicular directions of the applied magnetic field, which suppresses tensorial shifts such as the quadrupole shift~\cite{Itano2000}. While we use the first order Zeeman-insensitive $m_F=0 \rightarrow m_F=0$ transitions as the basis for both optical clocks, we periodically probe the $m_F=0\rightarrow m_F=2$ component of the E2 transition to determine the magnetic field strength for all three settings. The result is used to calculate a time-resolved correction for the second-order Zeeman shift on the E2 $m_F=0 \rightarrow m_F=0$ transition. Similarly, we calculate a dynamical correction of the shift due to black-body radiation based on measurements with resistive temperature sensors. Since most other parameters are kept constant during clock operation, the reproducibility of $\nu_\textrm{E2}$ is expected to be much smaller than its uncertainty. The dynamic correction of the second-order Zeeman shift leads to a reproducibility of $<1\times 10^{-18}$ for this shift. Since the quadrupole shift in our system remains constant over time at the $1\times10^{-16}$ level, and we suppress this shift by at least a factor of 100 with our averaging scheme \cite{Lange2020}, we estimate the reproducibility of the remaining quadrupole shift to $1\times 10^{-18}$. With these two changes we estimate the total fractional reproducibility of $\nu_\textrm{E2}$ in our system as $4\times 10^{-18}$, a significant improvement over the previously reported value of $16\e{-18}$ \cite{Lange2021}. 

In total, we evaluate about 235 days of measurement data of the ratio $\nu_\textrm{E3}/\nu_\textrm{E2}$, which were accumulated over a period of about 26 months, between MJD 59097 (September 5, 2020) and MJD 59900 (November 11, 2022). Combining this new data with that in \cite{Lange2021}, we find a linear drift of $\frac{1}{\nu_\textrm{E3}/\nu_\textrm{E2}}\frac{\textrm{d}(\nu_\textrm{E3}/\nu_\textrm{E2})}{\textrm{d}t}=-1.2(1.8)\times10^{-18}/\textrm{yr}$ in our measurements of the frequency ratio $\nu_\textrm{E3}/\nu_\textrm{E2}$, corresponding to a drift of the fine structure constant of
$\frac{1}{\alpha}\frac{\textrm{d}\alpha}{\textrm{d}t}=1.8(2.5)\times10^{-19}/\textrm{yr}$. For a possible coupling to the gravitational potential of the sun, $\Phi$, the best fit to all data gives $ \frac{c^2}{\alpha}\frac{\textrm{d}\alpha}{\textrm{d}\Phi}=-2.4(3.0)\times 10^{-9}$. Both values are compatible with zero and improve the uncertainty of the previous limits~\cite{Lange2021} by about a factor of four.

The fractional instability of the $\nu_\textrm{E3}/\nu_\textrm{E2}$ measurements is limited to $1.0\times 10^{-14}/\sqrt{\tau (\mathrm{s})}$ by the short interrogation time and limited duty cycle of the E2 interrogation. In order to overcome this limitation and investigate the full potential of the clock operating on the E3 transition, we complement the long-term measurements described above with a short measurement campaign comparing the \textsuperscript{171}Yb\textsuperscript{+} E3 single-ion clock to a \textsuperscript{87}Sr lattice clock in the spring of 2022. This frequency ratio features a similar sensitivity to variations of $\alpha$, but improved measurement stability compared to our measurements of $\nu_\textrm{E3}/\nu_\textrm{E2}$. 

As described in \cite{Falke2014, Schwarz2020}, the laser of the strontium clock is stabilized to the average frequency of the $m_F= \pm 9/2$, $\Delta m_F = 0$ transitions in $^{87}$Sr, which is free of the linear Zeeman shift. 
Rabi interrogation of typically 650~ms leads to a clock instability of below $2 \times 10^{-16}/\sqrt{\tau(\mathrm{s})}$ \cite{Schwarz2020}.
Unlike in previous publications, we used a new physics package, Sr3, that features a vertically oriented optical lattice to suppress tunneling between the lattice sites and an in-vacuum radiation shield that ensures a highly homogeneous thermal environment.
These and other improvements, including state preparation by sideband cooling in the lattice, lead to a systematic uncertainty of $3\times 10^{-18}$ of the new apparatus.
Further details will be given a subsequent publication \cite{Sr3_2023}.
The measurement setup used for the comparison is described in \cite{Dorscher2021}. In total, we evaluate about 343 hours of measurement data of $\nu_\textrm{E3}/\nu_\textrm{Sr}$ taken over a period of about 41 days.\\
\begin{figure}
    \centering
    \includegraphics[width = \columnwidth]{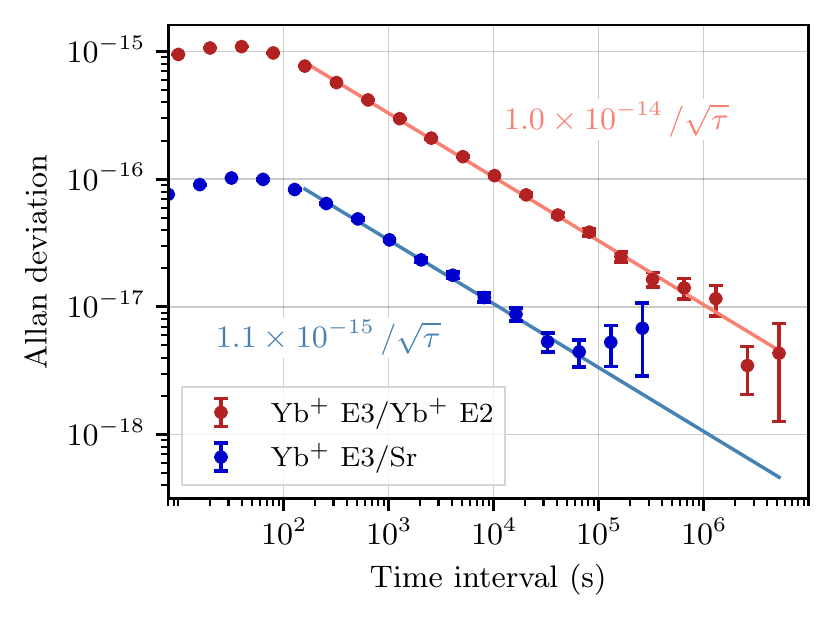}
    \caption{Instability of both frequency ratio measurements as characterized by the Allan deviation. The solid lines are fits to the Allan deviation with the $1/\sqrt{\tau}$ scaling expected for white noise.}
    \label{fig:allan}
\end{figure}
 The $\nu_\textrm{E3}/\nu_\textrm{Sr}$ measurement features a  fractional instability of $1.1\e{-15}\,/\sqrt{\tau (\mathrm{s})}$, limited by the quantum projection noise of the single ion clock. The instabilities of both frequency ratio measurements as characterized by the Allan deviation are shown in \autoref{fig:allan}. Deviation from the expected white noise behaviour is visible for $\nu_\textrm{E3}/\nu_\textrm{Sr}$ for averaging times above about $10^5\,$s. For the data analysis presented below, we model the noise in our measurements based on the Allan deviation: For the measurements of $\nu_\textrm{E3}/\nu_\textrm{E2}$ we assume pure white noise. For $\nu_\textrm{E3}/\nu_\textrm{Sr}$, we  additionally add random walk noise ($1/\omega^2$ power spectrum) with an amplitude of $9\times10^{-18}$, reproducing the increase of the Allan deviation we observe at long averaging intervals.\\ 

To search for a coupling of UBDM to photons, we look for sinusoidal modulations of the form
\begin{equation}
    S_\omega\,\sin(\omega t)+C_\omega\,\cos(\omega t)
    \label{eq:sin}
\end{equation}
in our measurement data of both frequency ratios, following closely the approach detailed in \cite{VanTilburg2015,Hees2016}. We are interested in the oscillation amplitude $A_\omega\equiv\sqrt{S_\omega^2+C_\omega^2}$ for a range of  angular frequencies $\omega$. To handle the gapped experimental data, we first estimate the power at different frequencies using the Lomb-Scargle formalism, a method for the spectral analysis of unevenly sampled data \cite{Scargle1982, VanderPlas2018}. This approach is equivalent to fitting the model \autoref{eq:sin} explicitly to the data for each frequency and constructing a periodogram from the $\chi^2$ goodness of fit:
\begin{equation}
   P_\omega=\frac{1}{2}\left[\chi_0^2-\chi^2(\omega)\right]\,\textrm{,}
\end{equation}
where $\chi_0^2$ is obtained from the non-varying reference model ($S\equiv C\equiv 0$). Using the Lomb-Scargle formalism as implemented in the astropy python package \cite{VanderPlas2015} instead of standard fitting routines speeds up the computation significantly. For evenly sampled data, the periodogram $P_\omega$ reduces to that obtained from a standard discrete Fourier transform. We extract the modulation amplitude from the periodogram via 
\begin{equation}
    A_\omega=\sqrt{4\,P_\omega/N_0}\,\textrm{,}
\end{equation}
with $N_0$ the number of datapoints. The highest frequency in our analysis is $0.005\,$Hz,  limited by the servo time of the optical clocks: For shorter times the laser frequencies are not yet fully determined by the atomic reference, and their stabilization to the same ultrastable optical cavity leads to common-mode rejection of signals. For a dataset where $T$ is the time between the beginning of the first measurement and the end of the last measurement, the Lomb-Scargle statistics are not valid for frequencies around $1/T$ and below, i.e. when signals no longer complete a whole oscillation cycle within the range of the data. In this case, we can still constrain the oscillation amplitude using a standard least-square fit of \autoref{eq:sin}, additionally allowing a constant offset for each frequency. Here, the limit on the amplitude increases with $1/\omega^2$, since being close to an antinode of the oscillation cannot be excluded. We check the agreement of both methods in a frequency range around $1/T$ and verify that our analysis reproduces the amplitudes of known reference signals.  

\begin{figure}
    \centering
    \includegraphics[width = \columnwidth]{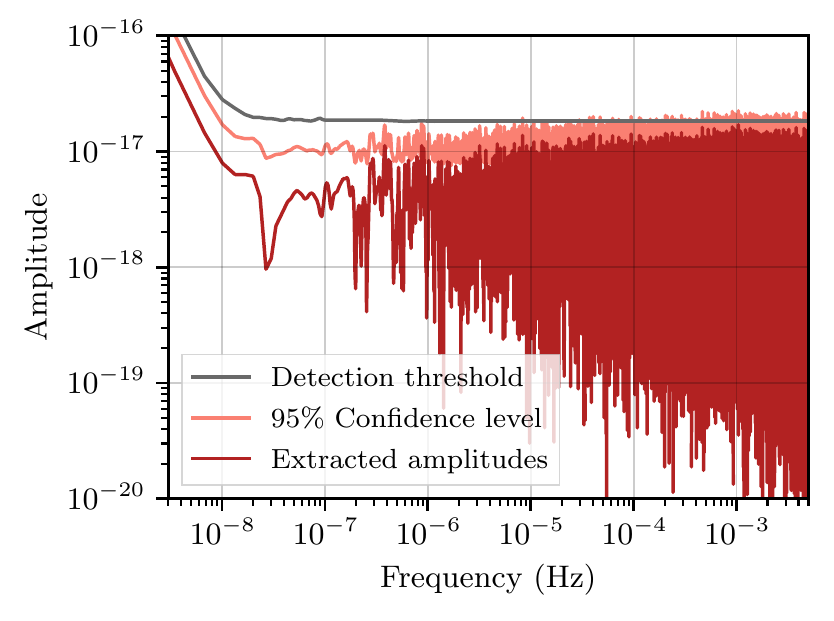}
    \caption{Amplitude spectrum of the $\nu_\textrm{E3}/\nu_\textrm{E2}$ measurement data (dark red) with upper 95\% confidence level (light pink) and 5\% detection threshold (gray)  (see text for details).}
    \label{fig:LombScargleE3E2}
%\end{figure}

%\begin{figure}[h!]
    %\centering
    \includegraphics[width = \columnwidth]{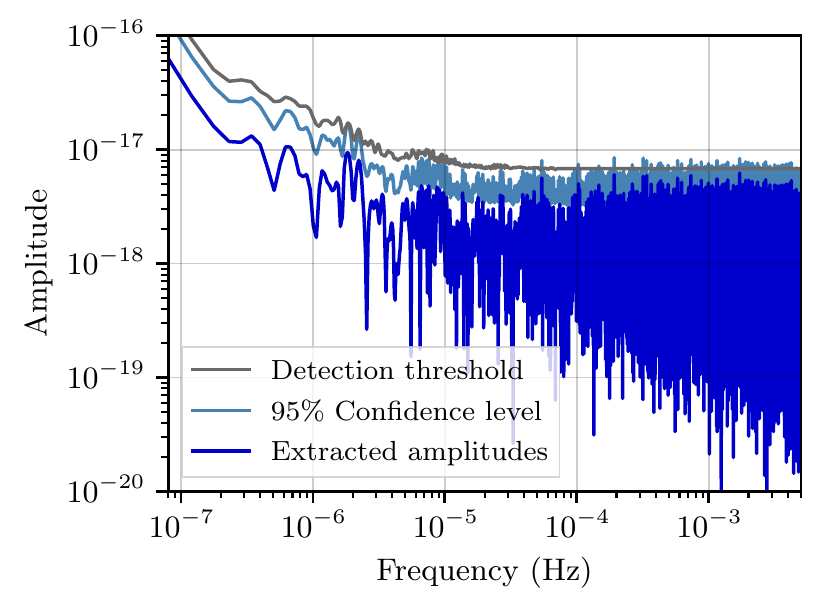}
    \caption{Amplitude spectrum of the $\nu_\textrm{E3}/\nu_\textrm{Sr}$ measurement data (dark blue) with upper 95\% confidence level (light blue) and 5\% detection threshold (gray)  (see text for details).}
    \label{fig:LombScargleE3Sr}
\end{figure}

We set an upper bound at $95\%$ confidence on the oscillation amplitude at each frequency by assuming the extracted amplitude is a real signal and generating 1000 datasets by adding a randomly generated offset to each datapoint according to our respective noise model. The random walk noise component is generated on a continuous dataset first, which is then gapped in accordance with the timestamps of our measurement. These datasets are then evaluated as described above, and taking the 95th percentile of the resulting distribution yields the upper 95\% confidence level \cite{VanTilburg2015,Hees2016,Kennedy2020}. In order to assess the statistical significance of any peaks in our spectrum, we additionally estimate a 5\% detection threshold. When finding a peak above this threshold, the probability of falsely assuming it to be a real signal would be less than 5\%. Our estimate of this detection threshold is based on Monte-Carlo (MC) sampling of noise (see Supplemental Material \cite{FilzinSuppl}) and converges with the analytic description in \cite{Scargle1982} for large frequencies.
%\footnote{See Supplemental Material at [URL will be inserted by publisher] for data that support the findings of this work.}.

The results of this analysis for the $\nu_\textrm{E3}/\nu_\textrm{E2}$ measurement data are presented in \autoref{fig:LombScargleE3E2}. Our 95\% confidence levels yield largely frequency-independent constraints on amplitude modulations below $2\times10^{-17}$ for frequencies $>1/T$. The results for the corresponding analysis of the $\nu_\textrm{E3}/\nu_\textrm{Sr}$ data are shown in \autoref{fig:LombScargleE3Sr}. The $\nu_\textrm{E3}/\nu_\textrm{Sr}$ data offers almost a factor 3 stricter amplitude constraints than the $\nu_\textrm{E3}/\nu_\textrm{E2}$ data for frequencies above the mid-$10^{-6}$ Hz range. The visible increase in the best-fit amplitudes for smaller frequencies is explained by our noise model and corresponds to the deviation from white noise visible in \autoref{fig:allan}. For even smaller frequencies $<1/T\approx 3\times 10^{-7}\,$Hz, we additionally see the expected $1/\omega^2$ scaling. 

We find no amplitudes exceeding the respective detection threshold in either dataset and conclude that there is no statistically significant sinusoidal modulation present in either of our measurements. Thus, our data does not indicate an UBDM-photon coupling given the constraints of the known measurement noise. In the following, we use the extracted upper 95\% confidence levels on sinusoidal modulations in our measurements to derive limits on such modulations of $\alpha$ and thus the coupling $d_e$ of UBDM to photons.

\begin{figure}
    \centering
    \includegraphics[width = \columnwidth]{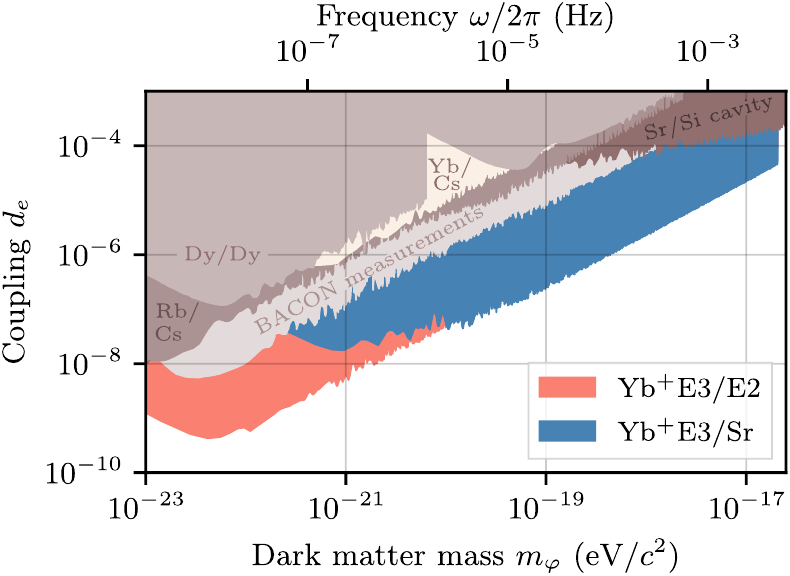}
    \caption{Exclusion plot for the coupling of ultralight bosonic dark matter to photons. The new limits deduced from our measurements of $\nu_\textrm{E3}/\nu_\textrm{E2}$ ($\nu_\textrm{E3}/\nu_\textrm{Sr}$) are shown in light pink (light blue) based on the 95\% confidence levels in \autoref{fig:LombScargleE3E2} (\autoref{fig:LombScargleE3Sr}). Also included are previous atomic limits reproduced from the literature: Those based on the radiofrequency spectroscopy of different isotopes of dysprosium (Dy/Dy) \cite{VanTilburg2015}, the frequency comparison of microwave atomic clocks with rubidium and cesium (Rb/Cs) \cite{Hees2016}, the comparison of a strontium lattice clock and a silicon cavity (Sr/Si) \cite{Kennedy2020}, several optical clock comparisons from the Boulder atomic clock optical network (BACON) \cite{Beloy2021}, as well as the comparison of an Yb lattice clock with a Cs fountain (Yb/Cs) \cite{Kobayashi2022}.}
    \label{fig:limits}

\end{figure}

Assuming the field $\varphi$ with mass $m_{\varphi}$ comprises all of the dark matter, we can translate our limits on the amplitude $A_\omega$ of a sinusoidal modulation with frequency $\omega/2\pi$ in our data to limits on the absolute value of the scalar coupling $d_e$ by combining \crefrange{eq:alpha}{eq:sens}:
\begin{equation}
    \left| d_e\right |=\frac{\omega A_\omega}{k_\alpha}\sqrt{\frac{c^2}{8\pi G\rho_{DM}}}\,\textrm{,}
\end{equation}

with an ambient dark matter energy density of $\rho_{DM}\approx6.4\times 10^{-5}\,\textrm{J}/\textrm{m}^3\approx0.4\,\textrm{GeV}/\textrm{cm}^{3}$ \cite{Kimball2022}. The limits deduced from the $95\%$ confidence levels shown in \autoref{fig:LombScargleE3E2} and \autoref{fig:LombScargleE3Sr} are depicted in the exclusion plot \autoref{fig:limits}. We reproduce previous limits in this mass range from the literature for reference. The total time spanned by our measurements of $\nu_\textrm{E3}/\nu_\textrm{E2}$ is $T\approx2\,\textrm{years}$, while the shortest coherence time, corresponding to the largest investigated mass $m_{\varphi}\approx2\times10^{-17}\,\textrm{eV}$, is about 6 years. Since $T < \tau_\mathrm{coh}$, no corrections due to finite coherence are necessary. We incorporate the effect of stochastic fluctuations of the dark matter amplitude by re-scaling our limits with a factor of 3 as suggested in~\cite{Centers2021}. This factor has also been applied to all data shown in \autoref{fig:limits} that originally neglected stochastic fluctuations.

Our constraints of $\left|d_e\right|$ based on the measurements of $\nu_\textrm{E3}/\nu_\textrm{Sr}$ are about a factor of 3 more stringent than those based on $\nu_\textrm{E3}/\nu_\textrm{E2}$ for masses above $10^{-20}\,\textrm{eV}/c^2$. For smaller masses, the long-term  measurement of $\nu_\textrm{E3}/\nu_\textrm{E2}$ yields tighter limits. Our combined results yield more than an order of magnitude improvement over previous bounds for masses ranging from the mid-$10^{-23}$ to the mid-$10^{-18}\,\textrm{eV}/c^2$ range. 
 
We note that for masses below $\approx10^{-22}\,\textrm{eV}/c^2$, corresponding to a de Broglie-wavelength the size of a small galaxy, the assumption that the field of mass $m_\varphi$ makes up all of the dark matter needs to be relaxed \cite{Kimball2022}, which is not considered in any of the depicted limits.

In summary, we substantially improve constraints on the coupling of ultralight bosonic dark matter to photons utilizing the high sensitivity of the \textsuperscript{171}Yb\textsuperscript{+} E3 transition to variations of the fine structure constant in two optical clock comparisons. While the E3 transition offers the highest sensitivity to $\alpha$ variations in currently operational optical clocks, a potential future nuclear optical clock based on thorium \cite{Flambaum2006, Peik2021}, would offer much higher sensitivity, and could potentially investigate a coupling several orders of magnitude below the limits presented here. Certain species of highly charged ions could also offer improved sensitivity \cite{Berengut2010, Safronova2014}. On the other hand, even with the sensitivities employed in this work, improved searches can be conducted, for example with a yearslong frequency comparison between a clock based on the E3 transition and a highly stable partner clock that features a small sensitivity to $\alpha$-variations. Additionally, improved stability of the clock based on the E3 transition can be obtained by achieving longer laser-ion coherence times and/or interrogating multiple ions simultaneously. In terms of the investigated mass range, applying the high $\alpha$-sensitivity of the E3 transition to larger dark matter masses around $10^{-16}\,\textrm{eV}/c^2$ and beyond is of interest. To achieve this without being limited by the typical servo time constants in optical clocks, dynamical decoupling sequences, which offer increased sensitivity to variations at particular frequencies, could be employed~\cite{Kennedy2020,Aharony2021}.  \\

We thank Aur\'{e}lien Hees for helpful discussions, Jialiang Yu, Thomas Legero and Uwe Sterr for providing a stable laser oscillator, and Burghard Lipphardt for experimental support.
We acknowledge support by the project 20FUN01 TSCAC, which has received funding from the EMPIR programme co-financed by the Participating States and from the European Union's Horizon 2020 research and innovation programme, and by the Deutsche Forschungsgemeinschaft (DFG, German Research Foundation) under Germany's Excellence Strategy -- EXC-2123 QuantumFrontiers -- Project-ID 390837967 and SFB~1227 DQ-\textit{mat} -- Project-ID 274200144 -- within project B02.
This work was partially supported by the Max Planck--RIKEN--PTB Center for Time, Constants and Fundamental Symmetries.

\section*{SUPPLEMENTAL MATERIAL}
\subsection{Constraints on a linear temporal variation of the fine-structure constant and its coupling to gravity}\label{sec:lpi}
\begin{figure}

    \centering
    \includegraphics[width = \columnwidth]{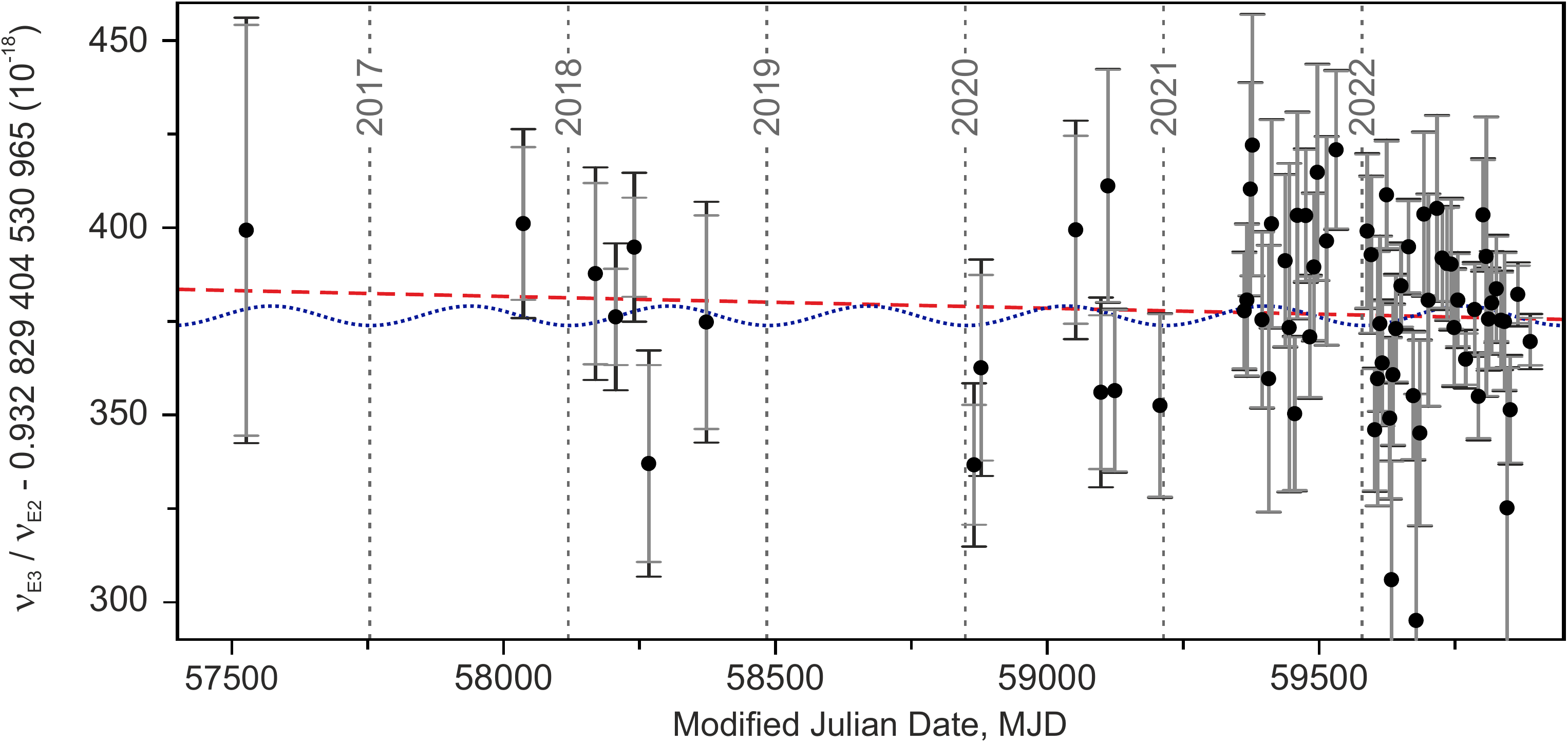}
    \caption{Ratio of the frequencies $\nu_{E3}$ and $\nu_{E2}$ of the E3 and the E2 transition in \textsuperscript{171}Yb\textsuperscript{+} measured between MJD 57527 (May 19, 2016) and MJD 59900 (November 11, 2022). The gray error bars show the statistical uncertainties. For the black error bars, a fractional uncertainty has been added in quadrature to take into account the reproducibility of systematic shifts over the measurement period. The dashed red line and the dotted blue line are fits to the data for searches for a linear temporal drift and a dependence on the gravitational potential, respectively.}
    \label{fig:alphavar}

\end{figure}

Previous optical frequency comparisons of $\nu_\textrm{E3}$ and $\nu_\textrm{E2}$ in our group found the strictest limits on a linear temporal variation of the fine-structure constant $\alpha$ and its coupling to the gravitational potential of the sun, which would lead to an oscillation of $\alpha$ with a period of one year \cite{Lange2021}. Including the new data presented in the main text in the same analysis allows us to further improve these limits. \Autoref{fig:alphavar} shows both the previously published measurement results of the frequency ratio $\nu_\textrm{E3}/\nu_\textrm{E2}$ between May 2016 (MJD 57527) and August 2020 (MJD 59081), as well as the new measurement data  starting from September 2020. The increased amount of data from 2021 and 2022 compared to earlier years is due to increased measurement activity as well as improvements in clock availability during measurements. The dashed red line and the dotted blue line are fits to all data of a linear temporal drift and a
sinusoidal modulation with a period of about 365 days respectively. Taking into account the statistical uncertainties and reproducibility, the fits result in a reduced $\chi^2$ of 1.2 and 1.1, respectively. We find a linear drift of
\begin{equation}
    \frac{1}{\alpha}\frac{\textrm{d}\alpha}{\textrm{d}t}=1.8(2.5)\times10^{-19}/\textrm{yr}\,\textrm{.}
\end{equation}
For a possible coupling to the gravitational potential $\Phi$ of the sun, the best fit to all data gives
\begin{equation}
    \frac{c^2}{\alpha}\frac{\textrm{d}\alpha}{\textrm{d}\Phi}=-2.4(3.0)\times 10^{-9}\,\textrm{.}
\end{equation}
Both values are compatible with zero and improve the previous best limits \cite{Lange2021} by about a factor of four in the uncertainty.

\subsection{Detection threshold}

In order to estimate a detection threshold for each frequency $\omega/2 \pi$, we need to account for the fact that extreme values become more likely as one considers a larger number of independent samples (look-elsewhere effect). We define the threshold $P_{\mathrm{th},\omega}$, following \cite{Scargle1982, Hees2016}, such that:
\begin{align}
\textrm{Prob}(P_\omega<P_{\mathrm{th},\omega })=(1-p_0)^{1/n_{\textrm{ind}}}
\end{align}
with the number of independent frequencies $n_\textrm{ind}$. In this case, if we find a value $P_\omega \geq P_{\mathrm{th},\omega}$ anywhere in the spectrum and interpret it as a detection, the probability of it being a false detection is less than $p_0=5\%$. We estimate the number of independent frequencies to be $n_\textrm{ind}\approx f_\mathrm{max} T$, where $f_\mathrm{max}=0.005$\,Hz, and obtain $n_\textrm{ind}\approx3.5\times10^{5}$ for the E3/E2 data and $n_\textrm{ind}\approx1.8\times10^{4}$ for the E3/Sr data. For $\omega/2\pi\gg 1/T$ and pure Gaussian white noise, the periodogram follows an exponential probability distribution, and the corresponding detection threshold is known analytically~\cite{Scargle1982}.

For smaller frequencies and different kinds of noise, where the probability distribution of the estimated power spectrum is not known, the detection limit can in principle be obtained directly from Monte-Carlo (MC) sampling of noise. However, this would require an extremely large number of samples, that is not computationally feasible. For the total evaluated measurement data of $\nu_\textrm{E3}/\nu_\textrm{E2}$, $0.95^{1/n_\textrm{ind}}\approx1-(2\times10^{-7})$, meaning that on the order of $10^7$ Monte-Carlo samples are necessary to obtain the detection threshold directly from the corresponding percentile of the probability distribution. A variety of approaches to this problem have been developed in different fields and for different statistics \cite{Gross2010, Cowan2011, Beaujean2018}. Here, we make use of the fact that an exponential distribution yields a good fit to the probability distribution of the estimated power at a given frequency, when allowing for a constant offset, and extract the threshold based on such a fit for a smaller number of MC samples. Convergence of the obtained results is at the few-percent-level already for $N=1000$ MC samples, and the quality of the fit to the results of the MC sampling is good throughout the parameter space considered, even for small frequencies and in the case of a random-walk noise component.

To determine the detection threshold, we thus fit the cumulative of an exponential distribution
\begin{align}
    1-\textrm{e}^{-a(P-P_0)}
\end{align}
to a histogram of the cumulative extracted power $P$, and determine the detection threshold based on the resulting fit parameters $a$ and $P_0$, then calculate the corresponding amplitude. We confirm the agreement of this method with the analytical detection threshold given in \cite{Scargle1982} for intermediate frequencies, and rely on the analytic result for frequencies $\gg 1/T$. In the case of $\nu_\textrm{E3}/\nu_\textrm{E2}$, the noise is well-described by white noise throughout, and since $1/T\approx1.4\e{-8}$, we rely on the analytic result for frequencies larger than $10^{-6}\,$Hz. For $\nu_\textrm{E3}/\nu_\textrm{Sr}$, we rely on the analytic result for frequencies larger than $4\times 10^{-5}\,$Hz, where the $1/\omega^2$ noise component is negligible and the measurement is dominated by white noise. The visible structure of the detection threshold at low frequencies results from the specific distribution of sampling gaps in our data.

\end{document}